\documentstyle[preprint,aps]{revtex}
\tightenlines

\begin{document}

\title{Security of Continuous Variable Quantum Cryptography}
\author{T.C.Ralph}
\address{Department of Physics, Centre for Laser
Science, \\ University of Queensland, \\ St Lucia, QLD 4072 Australia \\ 
Fax: +61 7 3365 1242  Telephone: +61 7 3365 3412 \\ E-mail: 
ralph@physics.uq.edu.au}
\maketitle


\begin{abstract}

We discuss a quantum key distribution scheme in which small phase and 
amplitude modulations of CW light beams 
carry the key information. The presence of EPR type correlations 
provides the quantum protection. We identify universal constraints on 
the level of shared information between the intended receiver (Bob) 
and any eavesdropper (Eve) and use this to make a general evaluation of 
security. We identify teleportation as an optimum eavesdropping 
technique.

\end{abstract}

\vspace{10 mm}

\section{Introduction}

The distribution of random number keys for cryptographic purposes can 
be made secure by using the fundamental properties of quantum 
mechanics to ensure that any interception of the key information can 
be detected \cite{wie83,ben92,eka92}. 
In particular the act of measurement in quantum mechanics inevitably 
disturbs the system. Further more, for single quanta such as a photon, 
simultaneous measurements of non-commuting variables are forbidden. 
By randomly encoding the information between non-commuting 
observables of a 
stream of single photons any eavesdropper (Eve) is forced to guess 
which observable to measure for each photon. On average, half the time Eve 
will guess wrong, revealing herself through the back action of the 
measurement to the sender (Alice) and receiver (Bob). There are a 
number of disadvantages in working with single photons, particularly 
the strong restrictions on data rates. Also it is of 
fundamental interest to quantum information research to investigate 
links between discrete variable, single photon phenomena and 
continuous variable, multi-photon effects. This has motivated a 
consideration of quantum cryptographic schemes using multi-photon light modes
\cite{ral99,hil00,rei00}.
 
The question of optimum protocols and 
eavesdropper strategies has been studied in detail for 
the single quanta case \cite{Fus96}, leading to general proofs 
of security for discrete systems. Up till now no such proofs of 
security have been made for continuous variable quantum cryptographic 
schemes. In this paper we introduce a scheme whose implementation and 
evaluation is sufficiently simple that a general proof of {\it 
minimum guaranteed security} can be derived. It is shown that in 
principle levels of security approaching those of single quanta 
schemes can be achieved.

The paper is 
laid out in the following way. In section II we review a 
coherent state scheme which provides limited security but none-the-less
serves to 
illustrate the basic concepts. In section III we derive in a general way 
the minimum disturbance Eve can make to the information Bob receives 
for a particular level of interception. This relationship is then 
used to make a general 
evaluation of the coherent scheme of section II. In section IV 
we extend our discussion 
to a scheme employing 2-mode squeezed states. We show that the 2-mode 
scheme can, in principle, provide high levels of guaranteed
security. In section V we make the interesting observation that 
teleportation represents an optimum eavesdropper strategy for this 
system and in 
section VI we conclude.

\section{Coherent State Quantum Cryptography}

In this section we will introduce the basic idea of our continuous 
variable cryptographic 
technique by reviewing a coherent state scheme \cite{ral99} which though 
not very secure, is illustrative. 
In both of our schemes we will consider encoding key information as small signals 
carried on the amplitude and and phase quadrature amplitudes of the 
beam. These are the analogues of position and momentum for a light 
mode and hence are continuous, conjugate variables. 
Although simultaneous measurements of these non-commuting observables 
can be made in various ways, for example splitting the 
beam on a 50:50 beamsplitter and then making homodyne measurements on 
each beam, the information that can be obtained is 
strictly limited by the generalized uncertainty principle for 
simultaneous measurements \cite{yam86,aut88}. If an ideal measurement of one 
quadrature amplitude produces a result with a signal to noise of
\begin{equation}
(S/N)^{\pm}={{V_{s}^{\pm}}\over{V_{n}^{\pm}}}
\end{equation}
then a simultaneous measurement of both quadratures cannot 
give a signal to noise result in excess of
\begin{equation}
(S/N)_{sim}^{\pm}=({{\eta^{\pm}V_{s}^{\pm}}\over{\eta^{\pm}
V_{n}^{\pm}+\eta^{\mp}V_{m}^{\pm}}}) S/N^{\pm}
\label{sn}
\end{equation}
Here $V_{s}^{\pm}$ and $V_{n}^{\pm}$ are, respectively, 
the signal and noise power of 
the amplitude ($+$) or phase ($-$) 
quadrature at a particular rf (radio frequency) with respect to the 
optical carrier. The quantum noise which is inevitably added when dividing the 
mode is $V_{m}^{\pm}$. The splitting ratio is $\eta^{\pm}$ and 
$\eta^{+}=1-\eta^{-}$ (e.g a 50:50 beamsplitter has 
$\eta^{+}=\eta^{-}=0.5$).
The spectral powers are normalized to the quantum 
noise limit (QNL) such that a coherent beam has $V_{n}^{\pm}=1$. Normally 
the partition noise will also be at this limit ($V_{m}^{\pm}=1$). For 
a classical light field, i.e. where $V_{n}^{\pm}>>1$ the penalty will 
be negligible. However for a coherent beam a halving of the signal to 
noise for both quadratures is unavoidable when the splitting ratio is 
a half. The 
Hartley-Shannon law \cite{har49} applies to Gaussian, additive noise, 
communication channels such as we will consider here. It shows, in general, 
that if information of a fixed bandwidth is being 
sent down a communication channel at a 
rate corresponding to the channel capacity and the signal to noise is 
reduced, then errors will inevitably appear at the receiver. Thus, 
under such conditions, any attempt by an eavesdropper to make 
simultaneous measurements will introduce errors into the transmission. 
In the following we will 
first examine what level of security is guaranteed by this uncertainty 
principle if a coherent state mode 
is used. We will then show that the level of security can in 
principle be made as strong as for the single quanta case
by using a special type of 
two-mode squeezed state. 

Consider the set up depicted in Fig.1. A possible protocol is 
as follows. Alice generates two 
independent random strings of numbers and encodes one on the phase 
quadrature, and the other on the amplitude 
quadrature of a bright coherent beam. Bob uses homodyne detection to 
detect either the amplitude or phase quadrature of the beam when he 
receives it. He swaps randomly which quadrature he detects. On a 
public line Bob then tells Alice at which quadrature he was looking, at 
any particular time. They pick some subset of Bob's data to be the test and 
the rest to be the key. For example, they may pick the amplitude 
quadrature as the test signal. They would then compare results for the times that 
Bob was looking at the amplitude quadrature. If Bob's results agreed 
with what Alice sent, to within some acceptable error rate, they would 
consider the transmission secure. They would then use the undisclosed 
phase quadrature signals, sent whilst Bob was observing the phase 
quadrature, as their key. By randomly swapping which quadrature is key 
and which is test throughout the data comparison an increased error 
rate on either quadrature will immediately be obvious.

Let us first quantify our results for some specific situations to 
illustrate the essential points. Consider the specific 
encoding scheme of binary pulse code modulation, in which the data is 
encoded as a train of 1 and 0 electrical pulses which are impressed on 
the optical beam at some rf using electro-optic modulators. 
The amplitude and phase signals are imposed at the same frequency with 
equal power. Let us now consider 
some specific strategies Eve could adopt (see 
Fig.2). Eve 
could guess which quadrature Bob is going to measure and measure it 
herself (Fig.2(a)). 
She could then reproduce the digital signal of that 
quadrature and impress it on another coherent beam which she would 
send on to Bob. She would learn nothing about the other quadrature 
through her measurement and would have to guess her own random string 
of numbers to place on it. When Eve guesses the right quadrature to 
measure Bob and Alice will 
be none the wiser, however, on average 50\% of the time Eve will guess 
wrong. Then Bob will receive a random string from Eve unrelated to the 
one sent by Alice. These will agree only 50\% of the time. 
Thus Bob and Alice would see a 25\% bit error rate in the test 
transmission if Eve was using this strategy. This is analogous to the 
result for single quanta schemes in which this type of strategy 
is the only available. Another single measurement 
strategy Eve could use is to do homodyne 
detection at a quadrature angle half-way between phase and amplitude. 
This fails because the signals become mixed. Thus Eve can tell when 
both signals are 0 or both are 1 but she cannot tell the 
difference between 1,0 and 0,1. This again leads to a 25\% bit error 
rate.

However, for bright beams it is possible to make simultaneous 
measurements of the quadratures, with the caveat that there will be 
some loss of information. So a second strategy that Eve could follow 
would be to split the beam in half, measure both quadratures and 
impose the information obtained on the respective quadratures of 
another coherent beam which she sends to Bob (Fig.2(b)). 
How well will this 
strategy work? Suppose Alice wishes to send the data to Bob with a 
bit error rate (BER) of about 1\%. For bandwidth limited transmission 
of binary pulse code modulation \cite{yar97}
\begin{equation}
BER={{1}\over{2}}erfc{{1}\over{2}}\sqrt{{{1}\over{2}}S/N}
\label{ber}
\end{equation}
Thus Alice must impose her data with a S/N of about 
13dB. For simultaneous measurements of a coherent state the signal to 
noise obtained is halved (see Eq.\ref{sn}). As a result, using 
Eq.\ref{ber}, we find the information Eve intercepts and subsequently 
passes on to Bob will only have a BER of 6\%. This is clearly a superior 
strategy and would be less easily detected. Further more Eve could 
adopt a third strategy of only intercepting a small amount of the 
beam and doing simultaneous detection on it (Fig.2(c)). For 
example, by intercepting 16\% of the beam, Eve could gain information 
about both quadratures with a BER of 25\% whilst Bob and Alice would 
observe only a small increase of their BER to 1.7\%. In other words Eve 
could obtain about the same amount of information about the key that she 
could obtain using the ``guessing'' strategy, whilst being very 
difficult to detect, especially in the presence of losses.

\section{Optimum Eavesdropper Strategy}

It is already clear from the preceding discussion that the coherent 
scheme does not provide good security. However our evaluation has 
been in terms of specific eavesdropper strategies. This is 
unsatisfactory for evaluating the security of more promising schemes. 
Instead we wish to be able to evaluate our schemes' security against 
attack from some theoretical, optimum 
eavesdropper strategy. Our approach is to identify the minimum 
disturbance allowed by quantum mechanics to the information Bob 
receives given a particular level of interception by Eve. We are then 
able to derive the minimum BER that Bob and Alice can find for a given 
BER in the information that Eve intercepts. We choose to couch our evaluation in 
terms of BER's because they represent an unambiguous, directly 
observable measure of the extent to which Eve can intercept 
information and the resulting corruption of Bob's information. 
Depending on the particular technique Eve uses Bob and 
Alice may be able to gain additional evidence for Eve's presence by 
comparing the absolute noise levels of the sent and received signals. 
This can only increase the security of the system. By considering a 
general limit on BER's we can find a minimum guaranteed security 
against eavesdropping regardless of the technique Eve employs.

A more general statement of the generalized uncertainty 
principle \cite{aut88} requires that for 
{\it any} simultaneous measurements of conjugate quadrature amplitudes
\begin{equation}
V_{M}^{+} V_{M}^{-}\ge 1
\label{gu}
\end{equation}
where $V_{M}^{\pm}$ are the measurement penalties for the 
amplitude ($+$) and phase ($-$) quadratures, normalized to the 
amplification gain between the system observables and the measuring 
apparatus. For example suppose an attempt to measure the amplitude 
quadrature variance of a system $V_{k}^{+}$ returned the result $G_{1} 
V_{k}^{+}+G_{2}V_{m}^{+}$ where $V_{m}^{+}$ represents noise. Then we 
would have $V_{M}^{+}=(G_{2}/G_{1})V_{m}^{+}$. Eq.\ref{sn} follows 
directly from  Eq.\ref{gu} for ideal simultaneous measurements. Let 
us investigate what general restrictions this places on the 
information that Eve can intercept and the subsequent corruption of 
Bob's signal. Firstly Eve's measurements will inevitably carry 
measurement penalties $V_{E}^{\pm}$ constrained by 
\begin{equation}
V_{E}^{+} V_{E}^{-}\ge 1
\label{eve}
\end{equation}
Now suppose Bob makes an ideal (no noise added) 
amplitude measurement 
on the beam he receives.  In order to satisfy Eq.\ref{gu} 
it must be true that the noise penalty 
carried on the amplitude quadrature of this beam $V_{B}^{+}$ due to 
Eve's intervention, 
is sufficiently large such that 
\begin{equation}
V_{B}^{+} V_{E}^{-}\ge 1
\label{bp}
\end{equation}
Similarly, 
Bob can also choose to make ideal measurements of the phase quadrature so we must 
also have 
\begin{equation}
V_{E}^{+} V_{B}^{-}\ge 1
\label{bm}
\end{equation}
Eqs.\ref{eve},\ref{bp},\ref{bm} are the quantum mechanical basis for 
our measure of minimum guaranteed security. They set strict limits on the 
minimum disturbance Eve can cause to Bob's information given a 
particular maximum quality of the information she receives. This 
applies regardless of the method she uses to eavesdrop. Given a 
particular encoding scheme, bandwidth and initial signal to noise, we 
can calculate minimum BER's from these results. 

Let us analyze the coherent state scheme using 
Eqs.\ref{eve},\ref{bp},\ref{bm}. In general the signal transfer 
coefficients, defined as the ratio of the output to input signal to 
noises are given by
\begin{eqnarray}
T_{E}^{+} & = & 
{{(S/N)_{eve}^{+}}\over{(S/N)_{in}^{+}}}=
{{V_{in}^{+}}\over{V_{in}^{+}+V_{E}^{+}}}\nonumber\\
T_{E}^{-} & = & 
{{(S/N)_{eve}^{-}}\over{(S/N)_{in}^{-}}}=
{{V_{in}^{-}}\over{V_{in}^{-}+V_{E}^{-}}}\nonumber\\
T_{B}^{+} & = & 
{{(S/N)_{bob}^{+}}\over{(S/N)_{in}^{+}}}=
{{V_{in}^{+}}\over{V_{in}^{+}+V_{B}^{+}}}\nonumber\\
T_{B}^{-} & = & 
{{(S/N)_{bob}^{-}}\over{(S/N)_{in}^{-}}}=
{{V_{in}^{-}}\over{V_{in}^{-}+V_{B}^{-}}}
\label{ts}
\end{eqnarray}
Substituting Eqs.\ref{ts} into Eqs.\ref{eve},\ref{bp},\ref{bm} and 
using the fact that $V_{in}^{\pm}=1$ we find
\begin{eqnarray}
T_{E}^{+}+T_{E}^{-} & \le & 1 \nonumber\\
T_{E}^{+}+T_{B}^{-} & \le & 1 \nonumber\\
T_{B}^{+}+T_{E}^{-} & \le & 1 
\label{ts2}
\end{eqnarray}
Eqs.\ref{ts2} clearly show that any attempt by Eve to get a good 
signal to noise on one quadrature (e.g. $T_{E}^{+}\to 1$) 
results not only in a 
poor signal to noise in her information of the other quadrature 
(e.g. $T_{E}^{-}\to 0$) but 
also a poor signal to noise for Bob on that quadrature (e.g. 
$T_{B}^{-}\to 0$), making her presence obvious. This is the general 
limit of the guessing strategy presented in the last section and leads 
to the same error rates.

Because of the symmetry of Bob's readout technique Eve's best 
approach is a symmetric attack on both quadratures. Eqs.\ref{ts2} then 
reduces to two equations
\begin{eqnarray}
2 T_{E}^{\pm} & \le & 1 \nonumber\\
T_{E}^{\pm}+T_{B}^{\pm} & \le & 1
\label{ts3}
\end{eqnarray}
If Eve extracts her maximum allowable signal to noise transfer, 
$T_{E}^{\pm}=0.5$, then ideally Bob suffers the same penalty 
$T_{B}^{\pm}=0.5$. This is the general limit of the second strategy of the 
previous section. The same reduction in Bob's signal to noise occurs 
as in the specific implementation (Fig.2(b)) thus this implementation can be 
identified as an optimum eavesdropper strategy for obtaining maximum 
simultaneous information about both quadratures. 

Eve's best strategy is to intercept only as much information as she 
can without being detected. The system will be secure if that level 
of information is negligible. Suppose, as in the last section, Eve 
only intercepts a signal transfer of $T_{E}^{\pm}=.08$. 
From Eq.\ref{ts3} this means Bob can receive at most a signal transfer of 
$T_{B}^{\pm}=.92$. This is greater than the result for the specific 
implementation shown in Fig.2(c), thus that implementation is not  
an optimum eavesdropper strategy. Using the optimum eavesdropper 
strategy the error rates for the specific 
encoding scheme discussed in the last section will be: if Eve 
intercepts information with a BER of 
25\%, then the minimum BER in Bob's information will be 1.4\%.

In this section we have derived a limit to the minimum 
back-action that Eve can produce on Bob's signal for a given quality 
of her intercepted signal. This limit is completely general, not 
dependent on the specific eavesdropping scheme employed. We have used 
our limits to investigate more generally the security of the coherent 
state scheme, coming to the same conclusion, i.e. that it is not very 
secure.

\section{Squeezed State Quantum Cryptography}

The preceding discussion has shown that a cryptographic scheme based 
on coherent light provides much less security than single quanta 
schemes. We now consider whether squeezed light can offer improved 
security. For example amplitude 
squeezed beams have the property $V_{n}^{+}<1<V_{n}^{-}$. 
Because the amplitude quadrature 
is sub-QNL greater degradation of S/N than the coherent case 
occurs in simultaneous 
measurements of amplitude signals (see Eq.\ref{sn}). Unfortunately the 
phase quadrature must be super-QNL, thus there is less degradation of 
S/N for phase signals. As a result the total security is in fact less 
than for a coherent beam. However in the following we will show that by
using two squeezed light beams, security comparable to that achieved 
with single quanta can be obtained. The following scheme is similar to 
that presented in our previous publication \cite{ral99} except that 
now only a single quantum limited beam is sent from Alice to Bob. 
This removes the possibility of Eve making a coherent attack, a 
loophole which (contrary to the claims in Ref.\cite{ral99})
was not adequately resolved in the previous protocol. 

The set-up is shown in Fig.3. Once again Alice encodes her number 
strings digitally, but now she impresses them on the amplitude 
quadratures of two, phase locked, 
amplitude squeezed beams, $a$ and $b$, one on each. A $\pi/2$ phase 
shift is imposed on beam $b$ and then they are mixed on a 
50:50 beamsplitter. The resulting output modes, $c$ and $d$, are given by
\begin{eqnarray}
c & = & \sqrt{{{1}\over{2}}}(a+i b)\nonumber\\
d & = & \sqrt{{{1}\over{2}}}(a-i b)
\end{eqnarray}
These beams are now in an entangled state which will exhibit Einstein, 
Podolsky, Rosen (EPR) type correlations \cite{ein35,ral98}. Negligible 
information about the  signals can be extracted from the beams 
individually because the large fluctuations of the anti-squeezed 
quadratures are now mixed with the signal carrying squeezed 
quadratures. One of the beams, say $c$, is 
transmitted to Bob. The other beam, $d$, 
Alice retains and uses homodyne detection 
to measure either its amplitude or phase fluctuations, with respect to a 
local oscillator in phase with the original beams $a$ and $b$. She 
randomly swaps which quadrature she measures, and stores the results. 
Bob, upon receiving beam $c$, also randomly chooses to measure either 
its amplitude or phase quadrature and stores his results. After the 
transmission is complete Alice sends the results of her measurements 
on beam $d$ to Bob on an open channel. About half the time Alice 
will have measured a different quadrature to Bob in a particular time window. 
Bob discards these results. The rest of the data corresponds to times 
when they both measured the same quadratures. If they both measured 
the amplitude quadratures of each beam Bob adds them together, in which case
he can obtain the power spectrum
\begin{eqnarray}
V^{+} & = & <|(\tilde c^{\dagger}+\tilde c)+(\tilde d^{\dagger}
+\tilde d)|^{2}>\nonumber\\
 & = & V_{s,a}+V_{n,a}^{+}
\end{eqnarray}
where the tilde indicate Fourier transforms. Thus he obtains the data string 
impressed on beam $a$, $V_{s,a}$, 
imposed on the sub-QNL noise floor of beam $a$, $V_{n,a}^{+}$. 
Alternatively if they both measured the phase quadratures of each 
beam, Bob subtracts them, in which case he can obtain the power spectrum
\begin{eqnarray}
V^{-} & = & <|(\tilde c^{\dagger}-\tilde c)-(\tilde 
d^{\dagger}-\tilde d)|^{2}>\nonumber\\
 & = & V_{s,b}+V_{n,b}^{+}
\end{eqnarray}
i.e. he obtains the data string impressed on beam $b$, $V_{s,b}$, 
imposed on the sub-QNL noise floor of beam $b$, $V_{n,b}^{+}$. Thus 
the signals lie on conjugate quadratures but 
{\it both} have sub-QNL noise floors. This is the hallmark of the EPR 
correlation \cite{ou92}. As for the coherent state case Alice and Bob 
now compare some sub-set of their shared data and check for errors. If 
the error rate is sufficiently low they deem their transmission 
secure and use the undisclosed sub-set of their data as their key.

Consider now eavesdropper strategies. Eve must intercept beam $c$ if 
she is to extract any useful information about the signals from the classical 
channel (containing Alice's measurements of beam $d$) sent later. 
She can adopt the guessing strategy by 
detecting a particular 
quadrature of beam $c$ and then using a similar apparatus to Alice's 
to re-send the beam and a corresponding classical channel later. 
As before she will only guess correctly what Bob will measure half the 
time thus introducing a BER of 25\%. Instead she may try 
simultaneous detection of both quadratures of beam $c$. As in the 
coherent case the noise she introduces into her own measurement 
($V_{E}^{\pm}$) and that she introduces into Bob's ($V_{B}^{\pm}$) are 
in general limited according to Eqs.\ref{eve},\ref{bp} and \ref{bm}. 
However now the consequences of these noise limits on the signal to 
noise transfers that Eve and Bob can obtain behave quite differently 
because the signals they are trying to extract lie on sub-QNL 
backgrounds. Eve's signal transfer coefficients are given by
\begin{eqnarray}
T_{E}^{+} & = & {{V_{n,a}^{+}}\over{V_{n,a}^{+}+0.5 V_{E}^{+}}}\nonumber\\
T_{E}^{-} & = & {{V_{n,b}^{+}}\over{V_{n,b}^{+}+0.5 V_{E}^{-}}}
\label{seve}
\end{eqnarray}
and similarly Bob's are
\begin{eqnarray}
T_{B}^{+} & = & {{V_{n,a}^{+}}\over{V_{n,a}^{+}+0.5 V_{B}^{+}}}\nonumber\\
T_{B}^{-} & = & {{V_{n,b}^{+}}\over{V_{n,b}^{+}+0.5 V_{B}^{-}}}
\label{sb}
\end{eqnarray}
For the squeezed noise floors the same ($V_{n,a}^{+}=V_{n,b}^{+}=V_{n}$) 
we find the signal transfers are restricted via
\begin{equation}
4 V_{n}^{2}({{1}\over{T_{E}^{+}}}-1)({{1}\over{T_{E}^{-}}}-1)\ge 1
\label{te}
\end{equation}
\begin{equation}
4 V_{n}^{2}({{1}\over{T_{E}^{+}}}-1)({{1}\over{T_{B}^{-}}}-1)\ge 1
\label{tb}
\end{equation}
\begin{equation}
4 V_{n}^{2}({{1}\over{T_{B}^{+}}}-1)({{1}\over{T_{E}^{-}}}-1)\ge 1
\label{tbb}
\end{equation}
It is straightforward to show that a 
symmetric attack on both quadratures is Eve's best strategy as it 
leads to a minimum disturbance in both her and Bob's measurements. 
Using this symmetry to simplify Eq.\ref{te} 
leads to the following general restriction on the 
signal transfer Eve can obtain:
\begin{equation}
T_{E}^{\pm} \le {{2 V_{n}}\over{2 V_{n}+1}}
\label{tes}
\end{equation}
Once the squeezing exceeds 3 dB ($V_{n}=0.5$) the signal to noise that 
Eve can obtain simultaneously is reduced below that for the coherent 
state scheme. In the limit of very strong squeezing ($V_{n}\to 0$) Eve 
can extract virtually no information simultaneously. Similarly Bob's signal 
transfer is restricted according to:
\begin{equation}
{{T_{E}^{\pm} T_{B}^{\pm}}\over{(1-T_{E}^{\pm})(1-T_{B}^{\pm})}} 
\le 4 V_{n}
\label{tbs}
\end{equation}
If squeezing is strong then almost any level of interception by Eve 
will result in very poor signal transfer to Bob. As a numerical 
example consider the specific encoding scheme of section 2 and suppose 
the squeezing is 13 dB ($V_{n}=0.05$). If Eve makes an ideal 
simultaneous measurement then both she and Bob will obtain a signal 
transfer of .09. 
As a result,assuming initial S/N of 13dB and using 
Eqs\ref{ber}, \ref{tes}, \ref{tbs} we find the information Eve intercepts and 
Bob receives will both have a minimum  BER of about 24\%. In other words, the 
security against an eavesdropper using simultaneous measurements 
is now on a par with the guessing strategy. Eve must reduce the 
amount of information she intercepts to virtually nothing before 
Bob's change in BER becomes negligible. The trade-off between Eve and 
Bob's BER's are shown graphically in Fig.4 for various levels of 
squeezing. Also shown is the corresponding trade-off for an ideal 
single quanta scheme. For high levels of squeezing the results are 
comparable.

In any realistic situation losses will be present. Losses tend in 
general to reduce security in quantum cryptographic schemes 
\cite{bar96}. Eve can take advantage of losses by setting up very close to Alice
and effectively disguising herself as loss. In such a situation her 
signal transfer remains as in Eq.\ref{tes} but Bob's changes to
\begin{equation}
({{1}\over{T_{E}^{\pm}}}-1)({{1}\over{T_{B}^{\pm}}}-1-
{{(1-\eta)}\over{2 \eta V_{n}}}) \ge {{1}\over{4 V_{n}}}
\end{equation}
where $\eta$ is the transmission efficiency. In Fig.5 some numerical 
examples including loss are given. Although these examples 
demonstrate some tolerance to loss for our 
continuous variable system it should be 
noted that single quanta schemes can tolerate much higher losses 
\cite{butt98} 
making them more practical from this point of view. 

\section{Teleportation as the Optimum Eavesdropper Strategy}

In this section we show that Eve can use continuous variable 
teleportation \cite{vai94,bra98,ral98} as an optimum eavesdropper 
strategy. Quantum teleportation uses shared entanglement to convert 
quantum information into classical information and then back again 
(see Fig.6). In 
particular continuous variable teleportation uses 2-mode squeezed light 
as its entanglement resource. In the limit of very strong squeezing no 
information about the teleported system can be extracted from the 
classical channel but a perfect reproduction of the quantum system 
can be retrieved. On the other hand with lower levels of squeezing 
some information about the system can be obtained from the the 
classical channel but at the expense of a less than perfect 
reproduction. We show in the following that under particular operating 
conditions the disturbance in the teleported state is precisely the 
minimum required by the generalized uncertainty principle, given the 
quality of information that can be extracted from the classical channel. 
Teleportation thus constitutes an optimum eavesdropper strategy.

Eves strategy would be to send the field she intercepts from Alice 
through a teleporter, adjusted such that she can read some 
information out of the classical channel, but still reconstruct the 
field sufficiently well such that Bob and Alice don't see a large BER.
The classical channel of a lossless continuous variable teleporter can 
be written \cite{ral99,ral98}
\begin{eqnarray}
F_{c} & = & K(f_{in}+j_{1}^{\dagger})\nonumber\\
 & = & K(f_{in}+\sqrt{G}v_{1}^{\dagger}+\sqrt{G-1}v_{2})
\end{eqnarray}
where $f_{in}$ is the annihilation operator of the input to the 
teleporter and $j_{1}=\sqrt{G}v_{1}+\sqrt{G-1}v_{2}^{\dagger}$ is the 
annihilation operator for one of the entangled beams. The $v_{i}$ are 
the vacuum mode inputs to the squeezers, $G$ is the parametric gain of 
the squeezers and $K>>1$ is the measurement amplification 
factor. Being a classical channel simultaneous measurements of both 
quadratures can be made without additional penalty thus immediately 
Eve's measurement penalty is
\begin{equation}
V_{E}^{\pm}=2 G-1
\end{equation}
For no squeezing ($G=1$) $V_{E}^{\pm}=1$, the minimum possible for 
simultaneous detection of both quadratures (see Eq.\ref{eve}). For 
large squeezing ($G>>1$) $V_{E}^{\pm}$ become very large and Eve can 
obtain little 
information from the classical channel.

The output of the teleporter is given by
\begin{eqnarray}
f_{out} & = & \lambda f_{in}+j_{1}^{\dagger}-j_{2}\nonumber\\
 & = & \lambda f_{in}+(\lambda \sqrt{G}-\sqrt{G-1})v_{1}^{\dagger}+
(\sqrt{G}-\lambda \sqrt{G-1})v_{2}
\end{eqnarray}
where $\lambda$ is the gain of the teleporter and $j_{2}=\sqrt{G}v_{2}+
\sqrt{G-1}v_{1}^{\dagger}$ is the 
annihilation operator for the other entangled beam. Thus Bob's measurement 
penalty for ideal measurements of either of the quadratures is
\begin{equation}
V_{B}^{\pm}= {{(\lambda \sqrt{G}-\sqrt{G-1})^{2}+(\sqrt{G}-\lambda 
\sqrt{G-1})^{2}}\over{\lambda^{2}}}
\end{equation}
If Eve operates the teleporter with gain \cite{note1}
\begin{equation}
\lambda_{opt}={{1+V_{sq}^{2}}\over{1-V_{sq}^{2}}}
\end{equation}
where $V_{sq}=(\sqrt{G}-\sqrt{G-1})^{2}$, then Bob's noise penalty is
\begin{equation}
V_{B}^{\pm}(\lambda_{opt})={{1}\over{2 G-1}}
\end{equation}
and so Eve causes the minimum allowable disturbance, i.e. 
$V_{E}^{\pm}V_{B}^{\pm}=1$.

\section{Conclusion}
 
We have derived quantum mechanical limits to the minimum disturbance that 
can be observed in the data shared by Alice and Bob in a continuous 
variable cryptographic scheme, in the presence of an eavesdropper using 
an optimum eavesdropper strategy. Using these limits we have shown, 
for the first time, a continuous variable scheme which can, in principle, 
have a guaranteed minimum security comparable with 
those of discrete, single quanta systems. In practice the high levels of 
squeezing and low levels of loss required will restrict the 
applicability of this particular scheme. Never-the-less we believe 
this work can be viewed as a simple, but rigorous starting point from which 
more practical schemes may be developed. Some promising lines for future 
inquiry are: investigating different encoding and transmission 
protocols; developing tighter security bounds and; investigating the 
use of higher order entanglement.  

We have shown that continuous variable teleportation is an optimum 
eavesdropper strategy for our system. This identifies 
continuous variable teleportation as the optimum method for extracting 
information from a quantum object whilst causing the least possible 
disturbance. 

Most generally this system is an example of a 
new quantum information technology based on continuous variable, 
multi-photon manipulations. Such technologies may herald a new 
approach to quantum information.

\begin{figure}
\caption{Schematic of coherent light cryptographic set-up. AM is an 
 amplitude modulator whilst PM is a phase modulator.}
\end{figure}

\begin{figure}
\caption{Schemata of three eavesdropper strategies. Only (a) 
 is available in single quanta schemes. }
\end{figure}

\begin{figure}
\caption{Schematic of squeezed light cryptographic set-up. Sqza and 
 sqzb are phase locked squeezed light sources. Rna and Rnb are 
 independent random number sources. Bs and pbs are non-polarizing and 
 polarizing beamsplitters respectively. Half-wave plates to rotate the 
 polarizations are indicated by $\lambda/2$ and optical amplification 
 by $A$. The $\pi/2$ phase shift 
 is also indicated. HD stands for homodyne detection system.}
\end{figure}

\begin{figure}
\caption{Minimum guaranteed security of squeezed light scheme. 
Minimum allowable BER's in the data of Bob and Eve are plotted for 
various levels of squeezing. The dotted line is for an ideal single 
quanta system. }
\end{figure}

\begin{figure}
\caption{Minimum guaranteed security of squeezed light scheme. 
Minimum allowable BER's in the data of Bob and Eve are plotted for 
various levels of transmission efficiency for 95\% squeezing. }
\end{figure}

\begin{figure}
\caption{Schematic of teleportation being used as an optimum 
eavesdropper strategy. }
\end{figure}

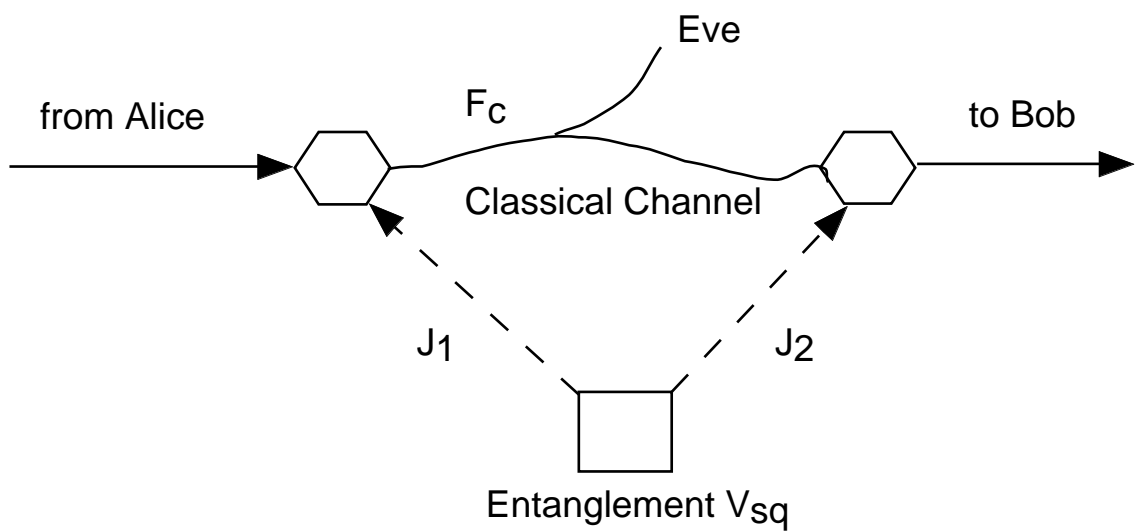
\end{document}